\documentclass[12pt]{article}
\usepackage[english]{babel}
\usepackage{ inputenc,amsmath,amsthm}
\usepackage{amsfonts}

\theoremstyle{definition}

\theoremstyle{remark}

\numberwithin{equation}{section}

\usepackage{amsmath}

\usepackage{bm}

\usepackage{setspace}

\title{Combining Data from Surveys and Related Sources}%
\author{Dexter Cahoy\\University of Houston-Downtown\\ J. Sedransk\\University of Maryland}%

\usepackage{natbib} 
\usepackage[plainpages=false,pdfpagelabels,hypertexnames,linktocpage, colorlinks, citecolor=red,linkcolor=blue]{hyperref}
\begin{document}

\maketitle

\begin{abstract}
To improve the precision of inferences and reduce costs there is considerable interest in combining data from several sources such as sample surveys and
administrative data. Appropriate methodology is required to ensure satisfactory inferences since the target populations and methods
for acquiring data may be quite different.
To provide improved inferences we use methodology that has a more general structure than the ones in current practice. We start with the case where the analyst has only summary
statistics from each of the sources. In our primary method, uncertain pooling, it is assumed that the analyst can regard one source, survey $r$, as the single best choice for inference. This method starts
with the data from survey $r$ and adds data from those other sources that are shown to form clusters that include survey $r$.
We also consider Dirichlet process mixtures, one of the most popular nonparametric Bayesian methods. We use analytical expressions and the results from numerical studies to show properties of the methodology.

\end{abstract}

\maketitle

\section{Introduction}

With substantially reduced response rates and limited budgets there has been an increased emphasis on efficient use of all of the information
available to the survey analyst. Specifically, one may be able to improve inferences by using results from several sample surveys and
related sources such as administrative records. The methodology that we use to combine information has more structure than the methods
currently used in survey sampling, so should lead to better inferences. Starting with the data from the survey that is the
best choice for inference, these data are augmented with other,
concordant, data. We use analytical expressions and the results from numerical studies to show properties of the methodology.

This research was motivated by a study \citep{has19} of health insurance coverage in Florida's counties where the authors noted very
different estimates from three surveys. Correspondingly, we could have estimates from a well established probability survey and two non-probability
surveys. In either case there is the question about how to make better inferences.

In the sequel we refer to the collection of studies as ``surveys," recognizing that these may be probability surveys, non-probability
surveys, administrative records and other sources. We consider the
case where the analyst has only a point estimate and associated standard error from each survey. This is common, as noted
in Section 7 of the review paper, \cite{lar17}. In the motivating example, and in many other cases, there are not any covariates that can be used to improve inferences. Our methodology extends to cases where the inferential objectives and models are more complex.

With survey estimates, $\{\hat{Y}_i : i = 1,...,L\}$, it is commonly assumed that  the $\hat{Y}_i$ are  independent

\begin{equation}\label{survey estimate}
\hat{Y}_i \sim N(\mu_i, V_i)
\end{equation}
where the $V_i$ are assumed to be known.

A common prior distribution expressing similarity among $\{\mu_1,..., \mu_L\}$ is
\begin{equation}\label{exchangeable prior}
\mu_i | \nu, \delta^2 \sim N(\nu, \delta^2)
\end{equation}
independently for each $i$, and $\nu$ and $\delta$ are assigned locally uniform prior distributions.

The resulting posterior expected value of $\mu_i$ is a convex combination of the estimate, $\hat{Y}_i$, for that survey, and
a weighted average of $\{\hat{Y}_1,...,\hat{Y}_L\}.$ The weakness of this approach is that the prior distribution in
(\ref{exchangeable prior}) assumes independent sampling of the $\mu_i$ from a common distribution.
The posterior mean is
\begin{equation}
E(\mu_i | \hat{Y}_1,..., \hat{Y}_L) = \lambda_i \hat{Y}_i + (1 - \lambda_i)\hat{Y}_w
\end{equation}
where $\lambda_i = \delta^2/(\delta^2 + V_i)$ and $\hat{Y}_w = \sum_{i=1}^L \lambda_i \hat{Y}_i/ \sum_{i=1}^L \lambda_i.$
This may lead to unsatisfactory inferences when, e.g., $\mu_1,..., \mu_b$ are each close to
$\mu*$ while $\mu_{b+1},...,\mu_L$ are each close to $\mu{**}$ and $\mu* << \mu{**}.$ Here,
estimation of $\mu_1$ would include, perhaps inappropriately, a large contribution from $\hat{Y}_{b+1},...,\hat{Y}_L$. The difficulty
is that the prior distribution is not sufficiently flexible. Here we use more flexible prior distributions that permit
the amount and nature of the pooling to be determined by the sample data.

The specification in (\ref{survey estimate}) and (\ref{exchangeable prior}) is common in meta-analyses and in situations where inferences for small
subpopulations and geographical areas are desired. For example, the U.S. Census Bureau uses such models (augmented by terms
to accommodate covariates) to make inference for U.S. county level poverty rates: see example 6.1.2 in \cite{rm15}. However, as just
noted, the assumptions in (\ref{survey estimate}) and (\ref{exchangeable prior}) of full exchangeability may not be appropriate, especially
for combining the information from $L$ surveys.

The theory for our principal method, uncertain pooling, was developed by \cite{mas92} and \cite{eas01} with further work in
\cite{eas99} and \cite{ys11}. To our knowledge this methodology has not been used in a survey sampling application:
the comprehensive review paper, \cite{lar17}, makes no reference to any technique similar to ours. We also modify
(\ref{survey estimate}) and (\ref{exchangeable prior}) by using a Dirichlet process mixture (DPM) as, e.g., employed by \cite{p17} for small area inference.

In this paper we describe the methodology for both uncertain pooling and DPM, and use them to analyze data from \cite{has19}. Then we modified these data
to exhibit properties of the methods. Finally, there is a simulation study to establish sampling properties. Clearly, the results from this
evaluation will apply equally if the three sources were, e.g., a well established probability survey and two non-probability surveys or other
choices.

We assume that the sample variances $\{V_1,..., V_L\}$ are specified. In our context none of the alternatives given in the literature for making
inferences for the $\{V_1,..., V_L\},$ all based on inferences for small areas, is fully satisfactory. In Section 4 we discuss this challenging problem of making inferences for the sample variances.

Both uncertain pooling and DPM have a more general structure than that in the common specification, (\ref{survey estimate}) and (\ref
{exchangeable prior}). This should lead to improved inferences. As seen in Section 2 the uncertain pooling model is the natural extension of (\ref{survey  estimate})
and (\ref{exchangeable prior}); i.e., the model in (\ref{survey estimate}) and (\ref{exchangeable prior}) is a special case.
Moreover, the output from uncertain pooling includes
the posterior probabilities associated with the possible clustering of the $L$ surveys.

Finally, note that we address only one of the many aspects of ``combining survey data," well summarized by \cite{lar17}.
Their Section 7, ``Hierarchical models for combining data sources," gives additional examples
where our methodology may be useful.

The methodology that we use is outlined in Section 2, and the results from our numerical studies are summarized in Section 3.
A brief summary and discussion are in Section 4.

\section{Methodology}

As in Section 1 assume that there are $L$ survey estimates, $\hat{Y}_1,..., \hat{Y}_L$, with
\begin{equation}\label{sample mean}
\hat{Y}_i \stackrel {ind}{\sim}N(\mu_i, V_i)
\end{equation}
where the $V_i$ are assumed to be known.

\subsection{Uncertain pooling}

The uncertain pooling method is based on \cite{mas92}  and \cite{eas01}. They showed
that a prior for $\mu=(\mu_1, \mu_2, \ldots, \mu_L)^t$  can be selected to reflect the beliefs that there are subsets of $\mu$ such that the $\mu_i$ in each subset are ``similar'', and that  there is uncertainty about the composition of such subsets of $\mu.$   Let $G$ be the total number of partitions of the set $\mathcal{L} =\{1,\ldots, L \},$ $g$ be a particular partition ($g= 1, \ldots, G$), $d(g)$ be the number of subsets of $\mathcal{L}$ in the $g$th partition $(1 \leq  d(g) \leq L)$, and $S_k(g)$ be the set of survey labels in subset $k (k= 1, ... , d(g)).$ For example, for $L = 3,$ there are $G = 5$ partitions: $\{g = 1\} \sim \{(123)\}, \{g = 2\} \sim \{(13), (2)\},
\{g = 3\} \sim \{(12), (3)\}, \{g = 4\} \sim \{(23), (1)\}, \{g = 5\} \sim \{(1), (2), (3)\}.$ Then, $S_1(g=2) = \{(13)\}$, $S_2(g = 2) = \{(2)\},
d_1(g =2) = 2$ and $d_2(g = 2) = 1.$

To specify a prior for $\mu,$  first condition on $g.$   \cite{mas92} and \cite{eas01}  assume that there is independence between  subsets, and within $S_k(g)$ the $\mu_i$  are independent with

\begin{equation}\label{first stage prior}
\mu_i |  \nu_k(g)  \sim N (\nu_k(g), \delta_k^2(g)),   \qquad i \in S_k(g) .
\end{equation}

Also, the $\nu_k(g)$ are mutually independent with
\begin{equation}\label {second stage prior}
\nu_k(g)  |  \theta_k(g)  \sim N (\theta_k(g), \gamma_k^2(g))
\end{equation}
where the $\theta_k(g)$ and the $\gamma^2_k(g)$ are hyperparameters. The definition in (2.3) is the first step in obtaining a reference prior for the $\nu_k(g)$, i.e., one that is dominated by the likelihood. This will include letting the $\gamma^2_k(g)
\rightarrow   \infty$, but is considerably more complicated, as described below. The $\delta^2_k(g)$ are also hyperparameters, to be assigned a prior distribution.

The formal definition in (\ref{second stage prior}) is included as the first step in obtaining a reference prior for the $\nu_k (g)$, i.e.,
one that is dominated by the likelihood, as described below in the evaluation of $f(g, \Delta^2 | y).$ Conditioning on the  $\delta^2_k(g)$ and $\gamma^2_k(g)$ (but suppressing them in our notation), and letting $\gamma_k^2(g) \to \infty$ leads to the following expected results for the posterior moments conditional on the partition $g.$
As discussed below, additional care is needed to obtain the posterior distribution of $g.$

Defining $y = (\hat{Y}_1,...,\hat{Y}_L)^t$, letting $\Delta^2 = \{\delta^2_k(g): k = 1,\ldots,d(g); g =
1,...,G\}$ and writing $\hat{\mu}_i = \hat{Y}_i$
\begin{equation}
\label{posterior expectation}
E(\mu_i | y,g, \Delta^2 ) = \{\lambda_i (g) \} \hat{\mu}_i + \{ 1-\lambda_i (g) \}  \hat{\mu}_k(g),  \qquad  i \in S_k(g)
\end{equation}
and
\begin{equation}
\label{posterior covariance}
cov (\mu_i, \mu_j | y, g, \Delta^2) = \left.
\begin{cases}
  \delta_k^2 (g) \lbrace 1-\lambda_i (g) \rbrace   + \frac{ \lbrace 1-\lambda_i (g)  \rbrace^2  \delta_k^2 (g) }{\sum_{i \in S_k(g)} \lambda_i (g)},   &   i=j; i,j \in S_k(g) \\
 \frac{ \lbrace 1-\lambda_i (g) \rbrace \lbrace 1-\lambda_j (g) \rbrace \delta_k^2 (g) }{\sum_{i \in S_k(g)} \lambda_i (g)},    &   i\neq j; i,j \in S_k(g) \\
 0, &      i  \in S_{k_1} (g), j  \in S_{k_2} (g), k_1 \neq k_2, \\
 \end{cases}
\right.
\end{equation}
where
\begin{equation}\label{lambda}
\lambda_i(g) =    \frac{\delta^2_k(g) }{  \delta^2_k(g)  + V_i },  \;  \hat{\mu}_k (g) = \frac{\sum_{j \in S_k(g)} \lambda_j(g) \hat{\mu}_j}{ \sum_{j \in S_k(g)} \lambda_j(g)}.  \;
\end{equation}

Note that $E(\mu_i | y, g, \Delta^2)$ has the familiar form of a weighted average of $\hat{\mu}_i$ and $\hat{\mu}_k(g)$, but, here, $\hat{\mu}_k(g)$ is restricted to the surveys in $S_k(g).$

Assuming the basic model in (\ref{survey estimate}) and (\ref {exchangeable prior}) corresponds, here, to the ``pool-all" partition, $\{g=1\}$,
where all of the $L$ surveys comprise a single cluster. Thus, for $\{g=1\}$ the moments in (\ref{posterior expectation}), (\ref{posterior
covariance}) and (\ref{lambda}) are those that would be obtained by an analysis using (\ref{survey estimate}) and (\ref {exchangeable prior}).
An analysis based on (\ref{survey estimate}) and (\ref {exchangeable prior}) is a special case of one based on the uncertain pooling
specification.

Inference about $\mu$ includes uncertainty about the value of $g,$ i.e.,
\begin{equation}\label{posterior mu}
f(\mu | y) = \int f(\mu | y, g, \Delta^2)f(g, \Delta^2 | y)dg d \Delta^2
\end{equation}
where the notation is simplified by using integration rather than summation for $g.$ Using the ``most likely" partition $g^*$ (i.e., $p(g^*|y) \ge p(g|y) : g = 1,...,G)$ to make inference would understate the overall precision.

To evaluate (\ref{posterior mu}) we need $f(g, \Delta^2 | y).$ However, when evaluating $f(g | \Delta^2, y)$ one must be careful about
specifying the rate at which the $\gamma^2_k(g)  \to \infty$: a natural choice leads to an expression for $f(g | \Delta^2, y)$
that is not invariant to changes in the scale of $Y$; see Section 4 of \cite{mas92}. \cite{mas92} provided a solution by using an empirical
Bayes argument. Here we use a fully Bayesian alternative, described in Section 5 of \cite{eas01}. It postulates little prior, relative
to sample, information about the $\nu_k(g)$, and is invariant to changes in the scale of $Y$.
Let $\nu(g) = (\nu_1(g), ..., \nu_{d(g)}(g))^t$
and $K\{f_1(\nu(g)), f_2(\nu(g) | y)\}$ be the Kullback-Leibler information about $\nu(g)$.
With prior $f(g, \Delta^2) = f(g)f(\Delta^2)$ and letting the $\gamma_k^2(g) \to \infty$ subject to $K\{f_1(\nu(g)), f_2(\nu(g) | y)\}$ = constant,
\begin{align}\label{posterior g, delta}
f(g, \Delta^2 |y)  \propto & \; f(\Delta^2) f(g)\exp \left\lbrace \frac{-d(g)}{2} \right\rbrace \prod \limits_{k=1}^{d(g)} \prod \limits_{i \in S_k(g)}  \lbrace 1-\lambda_i (g) \rbrace^{1/2} \nonumber \\
&\times \exp \left[ -\frac{1}{2} \sum\limits_{k=1}^{d(g)}  \sum\limits_{i \in S_k(g)} \left\lbrace \frac{\lambda_i(g)}{\delta^2_k(g)}  \right\rbrace \lbrace \hat{\mu}_i - \hat{\mu}_k(g) \rbrace^2  \right].
\end{align}
The term in the exponent,
\begin{equation}
 Q\{d(g)\} = \sum\limits_{k=1}^{d(g)}  \sum\limits_{i \in S_k(g)} \left\lbrace \frac{\lambda_i(g)}{\delta^2_k(g)}  \right\rbrace \lbrace \hat{\mu}_i - \hat{\mu}_k(g) \rbrace^2,
\end{equation}
is likely to decrease as $d(g)$ increases, for example for a new partition of $\bigcup_{k=1}^{d(g)} S_k(g)$ obtained by creating subsets
of the existing $\{S_k(g)\}$. Since $f(g, \Delta^2 | y)$  increases as $Q\{d(g)\}$ decreases, it is helpful to have the
second term, $\text{exp}\{-d(g)/2\}$, that penalizes partitions with larger values of $d(g)$.

For our analysis we take $\delta^2_k(g) = \delta^2$ and write $\lambda_i(g) = \delta^2/(\delta^2 + V_i).$
Inference for $\mu$ is made using (\ref{posterior mu}) and (\ref{posterior g, delta}) with
\begin{equation} \label{posterior mean}
\mu | y, g, \delta^2 \sim N(E(\mu | y, g, \delta^2), V(\mu | y, g, \delta^2))
\end{equation}
where the conditional posterior moments of $\mu$ are given in (\ref {posterior expectation}) and (\ref{posterior covariance}).

We assume that $f(g)$ is constant, i.e., that all partitions are equally likely, a priori, and take the Inverse Beta prior for
$\delta^2$, i.e.,

\begin{equation}\label{InvBeta}
f(\delta^2) \propto 1/(1+ \delta^2)\sqrt{\delta^2}, \quad 0 < \delta^2 < \infty.
\end{equation}

Inference for $\mu$ is made using (\ref{posterior mu}). To start, evaluate the right side of (\ref{posterior g, delta}) for
\begin{equation}
\{g: g = 1, . . ., G; \quad  R \; \text{grid points for} \; \delta^2 \;  \},
\end{equation}
then
standardize by dividing the individual terms in the grid by their sum. This provides an approximation for
$f(g, \delta^2 | y).$
Then select a random sample of size $B$ from the $RG$ normalized values of $f(g, \delta^2 | y).$ For each selection, $(g_{*},\delta^2_{*}),$
sample $\mu$
from  $f(\mu | y, g_{*}, \delta^2_{*}).$
Here, we generated $B = 5,000$ values of $\mu$.
Finally, note that approximations for the marginal posterior distributions, i.e., $f(g | y)$ and $f(\delta^2 | y),$ can be obtained
directly from the grid approximation of $f(g, \delta^2 | y)$.

Assuming that survey $r$ is the single best choice for inference we can consider the posterior distribution corresponding to
survey $r$ to be the object of
inference. In a common contemporary application there will be data from a well established probability survey (the single best
choice) and data from other sources such as non-probability surveys, administrative records, etc. In other settings it is likely
that there will be a preference for one of the surveys.

Then, using the posterior expected value for illustration,
\begin{equation}
E(\mu_r | y) = E_{g, \delta^2 | y}E(\mu_r | y, g, \delta^2)
\end{equation}
where $E(\mu_r | y, g, \delta^2)$ is defined in (\ref{posterior expectation}). Thus, inference for $\mu_r$ is a function of $\hat{\mu}_r$
together with data from the other $L - 1$ studies as determined by the form of (\ref{posterior expectation}), and, critically, by the likelihood
associated with the set of subsets, $S_k(g)$, containing study $r.$ See Evans and Sedransk (2001) for additional details and an application
to a notable study of the effect of using aspirin by patients following a myocardial infarction.

The model given by \cite{ch14} has a superficial resemblance to the one in (\ref{sample mean}), (\ref{first stage prior}), and (\ref{second
stage prior}). Taking $x_i$ to be the scalar with value $1$, the model in (2.1) of \cite{ch14} is
\begin{equation}
\hat{Y}_i = \mu_i + e_i, \quad i = 1,..., L
\end{equation}
where $\mu_i = \xi + (1 - \delta_i)\nu_{1i} + \delta_i \nu_{2i} + e_i$
with $e_i , \delta_i, \nu_{1i}, \nu_{2i}$  independent, $p(\delta_i = 1 | p) = 1 - p, \nu_{1i} \sim N(0, A_1)$ and
$\nu_{2i} \sim N(0, A_2).$ Finally, $e_i \sim N(0, V_i)$ with $V_i$ known.
Thus, unlike the uncertain pooling method, there is only a single focal point, $\xi.$ This permits appropriate treatment
of outliers, but does not take advantage of possible clustering of the $\mu_i.$ This can also be seen in (2.4) of \cite{ch14}
where
\begin{equation}\label{Chakraborty}
E(\mu_i | \xi, A_1, A_2, p, y) = \hat{Y}_i - \kappa_i(\hat{Y}_i - \xi)
\end{equation}
and $\kappa_i$ is a function of $V_i, A_1, A_2, p(\delta_i = 0 | \xi, A_1, A_2, p, y)$
with $y = (\hat{Y}_1,...,\hat{Y}_L)^t.$
\cite{ch14} show that if survey $i$ is an outlier, $E(\mu_i | \xi, A_1, A_2, p, y) \approx \hat{Y}_i$, as desired. Now suppose that surveys $\{1, \ldots, b\}$ and
$\{b+1, \ldots, L\}$ form two distinct clusters with a very large separation between them. Then inference for $\mu_1,$ say, will not,
in general, use the data in an appropriate manner. In (\ref{Chakraborty}) there should be two values of $\xi$, i.e.,
corresponding to the two subsets. And appropriate use of information about subsets is the essence of the uncertain pooling method.


\subsection{Dirichlet process mixture}

An alternative to the uncertain pooling method is to use a Dirichlet process mixture (DPM), one of the most popular nonparametric
Bayesian methods. This methodology is presented in detail in Sections 2.1 and 2.2 of \cite{mqjh15}. For our analyses we have used
the R function DPmeta from the package DPpackage: see \cite{jhqmr11} for details. The model in DPmeta is
\begin{equation}\label{DPM first stage}
y_i | \theta_i \stackrel{iid}{\sim} f_{\theta_i}
\end{equation}and
\begin{equation}\label{DPM second stage}
\theta_i | H \stackrel{iid}{\sim} H
\end{equation}
with $H \sim DP(M, H_0)$.

In (\ref{DPM first stage}) and (\ref{DPM second stage})  $y_i = \hat{Y}_i, \theta_i = \mu_i,  f_{\theta_i}$ is the pdf of a $N(\mu_i,
V_i)$ random variable with $V_i$ fixed, and $H_0 = N(\eta, \tau^2)$.

The (independent) hyperparameters are
\[ 
M | a_0, b_0 \sim \text{Gamma} (a_0, b_0)
\] 
\[ 
\eta | \eta_b, S_b \sim N(\eta_b, S_b)
\] 
\begin{equation}
\tau^{-2} | \phi_1, \phi_2 \sim \text{Gamma} (\phi_1/2, \phi_2/2).
\end{equation}

\cite{p17} has proposed using a DPM of this nature for inference about small area parameters. As in Section 2.1 of our
paper \cite{p17} indicates the value of extending the typical random effects model (e.g., the well known Fay-Herriot model)
that assumes full exchangeability of the set of small area parameters.

The uncertain pooling method requires only that one specify a prior distribution for $g$ and $\delta^2$. By contrast DPmeta requires
substantial prior input, i.e., values for $a_0, b_0, \eta_b, S_b, \phi_1$ and $\phi_2.$ Without strong prior information we can't
make proper inferences for these quantities with only $L = 3$ surveys. So, we have omitted the specification
$M | a_0, b_0 \sim \text{Gamma} (a_0, b_0)$ and made inference for a selected set of values of $M$ as suggested by \cite{e94}. Also, we
replaced $\eta_b, S_b, \phi_1$ and $\phi_2$ with their maximum a posteriori probability estimates.

\section{Results}

\cite{has19} made inference for the proportion of adults without health insurance in each of the 67 Florida counties, and compared these estimates with those from two other sources. Some of the differences were striking, motivating us to consider methodology to make appropriate inferences in such cases. We use these data, and modifications of these data, to show the benefits of using the methodology outlined in Section 2. One source is the Small Area Health Insurance Estimates Program (SAHIE, hereafter survey 1). The SAHIE program uses point estimates from the American Community
Survey (ACS) together with administrative data such as Federal income tax returns and Medicaid/Children's Health Insurance Program (CHIP) participation rates. There is
detailed area level modelling. The principal ones are models of ACS estimates of the proportions in income groups and the proportions insured.
There are additional models such as ones modelling the number of persons enrolled in Medicaid or CHIP,
Supplementary Nutrition Assistance Program (SNAP) participation, and IRS tax
exemptions. For a full understanding of this program see the twenty-two page technical report, \cite{bls18}. We have added a non-technical
summary in the Appendix.

The analyses using both survey 2, denoted by HS, based on \cite{has19}, and survey 3 denoted by CDC (Centers for Disease Control and
Prevention) use unit level models based on 2010 data from the Behavioral Risk Factor Surveillance System (BRFSS), obtained through telephone interviews.
While the sample designs differ somewhat over states, the one in Florida was typical, i.e., a disproportionate stratified sample design. In Florida,
the set of telephone numbers were divided into two strata (high and medium density) that were sampled separately. In addition there was a stratification
by area codes, i.e., three geographic strata and a fourth stratum consisting of area codes with large estimated Hispanic populations.
For additional, general information see http://www.cdc.gov/brfss/annual\_data/annual\_2010.htm
while for technical details see \cite{pxw16} and \cite{has19}. Both use, essentially, the same covariates but the modeling in HS is more detailed. Moreover, \cite{pxw16}
give only point estimates, noting that standard errors were being developed. A further complication is that the CDC analysis is frequentist  while the SAHIE and HS analyses are Bayesian. Thus, we have an (empirical) Bayes posterior standard deviation for SAHIE, none for CDC and an estimated SE for HS obtained by taking the 95\% credible interval for a county proportion and dividing by 3.92.
While the limitations just noted preclude definitive conclusions from these data they illustrate the methodology. Moreover, conditions where standard errors are missing or unreliable are, at least, fairly common for non-probability samples, a focus of this paper.

The first set of analyses is based on the observed data. To show other properties of the methodology a second set of analyses is based on
modifications of these data. Finally, to show sampling properties there is a simulation study. Note that each of our analyses is based \emph{only} on data from a single county. Additional research is needed to permit inference using data from all of the sources and counties. See Section 4 for discussion.

\subsection{Data-based analyses}

Using the uncertain pooling methodology a summary of the results for Dixie county is presented in Table 1. These are typical of most of the county-based analyses that we
have done. Throughout, SE denotes the sample standard error. There are three panels, corresponding to choices of the CDC SE, taken equal to 0.5, 1.0, 2.0 times the HS SE. For each panel the column headings are the observed proportion,  posterior mean of the county proportion,
estimated SE of the observed proportion, posterior standard deviation and lower and upper bounds of the 95\% credible interval for the county proportion.  At the bottom of each panel there are the values of $p(g | y)$ with $p(g|y) \ge 0.001$ where $\{g =1\} \sim \{(123)\}, \{g =2\} \sim \{(13), (2)\}, \{g = 3\} \sim \{(12), (3)\}, \{g = 4\} \sim \{(2,3), (1)\}$ and $\{g = 5\} \sim \{(1), (2), (3)\},$ and summaries corresponding to $\{g = 1\},$ labelled ``pool-all".

We first analyze these data using the uncertain pooling methodology, then compare them with those from DPmeta.

A common way to summarize a set of sample proportions is to assume that the corresponding set of true proportions come from a common
source, i.e., $p(g = 1) = 1$. However, for each of the three cases in Table 1, $p(g = 1 | y) \le 0.001.$ Thus, there is very little support for
pooling all of the data from the three surveys. For further investigation of the effect of assuming a common source, assume $g=1.$ Then,
as in (\ref{survey estimate}) and (\ref{exchangeable prior}),

\begin{align}
\hat{Y}_i \stackrel{ind}{\sim} N(\mu_i, V_i) &  \nonumber \\
\mu_i \stackrel{iid}{\sim} N(\nu, \delta^2), & \quad  i=1,\ldots, L.
\end{align}
With a locally uniform prior on $\nu$ and the Inverse Beta prior on $\delta^2$ in (\ref{InvBeta})
\begin{equation}
f(\nu | y) = \int f(\nu | \delta^2, y)f(\delta^2 | y)d\delta^2
\end{equation}
where the posterior distribution of $\nu$ given $\delta^2$ is normal with
$E(\nu | \delta^2, y) = \frac{\sum_{i=1}^L \hat{Y}_i/(V_i + \delta^2)}{\sum_{i=1}^L 1/(V_i + \delta^2)}$ and
$Var(\nu | \delta^2, y) = (\sum_{i=1}^L (V_i + \delta^2)^{-1})^{-1}.$

For panel 1 in Table 1, $E(\nu | y) = 0.313, \quad SD(\nu | y) = 0.017$ and the 95\%
credible interval is $(0.290, 0.340)$. Inferences based on the posterior distribution of $\nu$ are not consistent with the notion that
any one of the three surveys is the nominal ``gold standard." If, for example, survey 1 is taken as the ``gold standard," the posterior mean of $\mu_1, 0.254,$
is substantially smaller than the posterior mean of $\nu, 0.313$. Moreover, $0.254$ is not included in the 95\% interval for $\nu, (0.290, 0.340).$ The conclusions from panels 2 and 3 are essentially the same. Finally, recall that $p(g = 1 | y) \le 0.001,$ indicating very little
support for pooling all of the data.

\begin{table}[h!t!b!p!]
\centering
\caption{Observed proportions, standard errors and posterior summaries from Dixie County, Florida using uncertain pooling.}

\setlength\tabcolsep{2pt}
\begin{tabular}{cccccc}
 \hline
 & \multicolumn{4}{c}{CDC SE = 0.5 $\times$ HS SE} &   \\
 Survey   &  ObsProp   &  PostMean    &  ObsSE   &  PostSD &  $95\%$ Cred Int  \\
1  &  0.254   & 0.254  &  0.014    & 0.014 &   (0.225, 0.283)  \\
2   &0.361  &  0.360  &  0.028   &   0.020 &  (0.317, 0.403) \\
3  & 0.359   &0.359   & 0.014    &   0.013  & (0.333, 0.385)\\
pool-all&  & 0.313 &  & 0.017 &  (0.290, 0.340) \\
\multicolumn{6}{l}{\scriptsize $P(g=3|y)=0.002; P(g=4|y)=0.621; P(g=5|y)=0.377$} \\
    &   &   & &              &    \\
     & \multicolumn{4}{c}{CDC SE =  HS SE}  &   \\
Survey   &  ObsProp   &  PostMean    &  ObsSE   &  PostSD &  $95\%$ Cred Int  \\
  1  &   0.254 & 0.254 &  0.014 & 0.014 &    (0.225, 0.283)   \\
  2  &   0.361  &  0.360 & 0.028 & 0.023  &     (0.313, 0.406)    \\
  3  &  0.359   & 0.359  &0.028  & 0.023 &   (0.312, 0.404)  \\
pool-all&  & 0.290 &  & 0.011 &  (0.268, 0.312) \\
\multicolumn{6}{l}{\scriptsize $P(g=2|y)=0.002; P(g=3|y)=0.002;  P(g=4|y)=0.619;P(g=5|y)=0.376$} \\
   &   &   & &              &    \\
     & \multicolumn{4}{c}{CDC SE =  2 $\times$ HS SE}  &   \\
Survey   &  ObsProp   &  PostMean    &  ObsSE   &  PostSD &  $95\%$ Cred Int  \\
  1  &   0.254 & 0.254 &  0.014 & 0.014 &    (0.226, 0.284)   \\
  2  &   0.361  &  0.360 & 0.028 & 0.026  &     (0.307, 0.412)    \\
  3  &  0.359   & 0.349  &0.056  & 0.048 &   (0.256, 0.303)  \\
pool-all&  & 0.279 &  & 0.012 &  (0.255, 0.305) \\
\multicolumn{6}{l}{\scriptsize $P(g=1|y)=0.001; P(g=2|y)=0.107; P(g=3|y)=0.002;  P(g=4|y)=0.554;P(g=5|y)=0.336$} \\
     \hline
\end{tabular}

\end{table}

\begin{table}[h!t!b!p!]
\centering
\caption{Observed proportions,  standard errors  and posterior summaries from Dixie County, Florida using DPmeta.}

\setlength\tabcolsep{4pt}
\begin{tabular}{cccccc}
 \hline
Survey & ObsProp &  PostMean  &  ObsSE &  PostSD  &   $95\%$ Cred Interval  \\
     &  \multicolumn{5}{c}{CDC SE =  0.5 $\times$ HS SE}   \\
1  &  0.254     &  0.254 &0.014    & 0.014 &     (0.227, 0.282) \\
2   &0.361     &   0.359  &0.028&     0.013 &   (0.334, 0.384)  \\
3  & 0.359      & 0.360 &0.014     &   0.012  &   (0.335, 0.384)\\
   & &  &   &  &   \\
     &  \multicolumn{5}{c}{CDC SE =  HS SE}  \\
  1  &   0.254 &  0.256 & 0.014&   0.016 &       (0.227, 0.290)    \\
  2  &   0.361  &   0.357 &0.028&   0.024  &          (0.291, 0.399)  \\
  3  &  0.359   &    0.357 &0.028&  0.025 &        (0.290, 0.399) \\
   & &  &   &  &    \\
   &    \multicolumn{5}{c}{CDC SE =  2 $\times$ HS SE}   \\
  1  &   0.254 &  0.264 & 0.014&   0.018 &        (0.230, 0.287)  \\
  2  &   0.361  &    0.332 &0.028&   0.044  &          (0.261, 0.406)    \\
  3  &  0.359   &   0.321  &0.056&    0.048  &      (0.249, 0.402)   \\
     \hline
\end{tabular}

\end{table}

In the following assume, for illustration, that one prefers the HS methodology. Then there may be substantial gains in precision (measured by the posterior standard deviation) by using the uncertain pooling methodology. The gain in precision is measured by comparing
the posterior standard deviation from the uncertain pooling methodology with that obtained by using only the data from the specific survey, here HS (survey 2). For the latter and a locally uniform prior for $\mu_2$, the posterior distribution of $\mu_2$ is normal with posterior mean equal to the observed proportion and posterior standard deviation equal to the estimated SE. If we take  CDC SE = $k$(HS SE) for $k = 0.5, 1, 2$, then the reductions in the posterior standard deviation for HS (survey 2), and corresponding to $k = 0.5, 1, 2$, are 29, 18 and 7\%. (For example, from panel 1
of Table 1, i.e., $k = 0.5,$ the percent reduction in the posterior SD for HS is 100(0.028 - 0.020)/0.028)\% = 29\%. Note that the relatively small SEs for each of the surveys means that the ``all singletons" partition, i.e., $\{g =5\}$, has a relatively large posterior probability (about 0.38). The corresponding reductions in the posterior standard deviation (uncertain pooling vs. no pooling) for CDC (survey 3) are 7, 18 and 14\%.

As noted in Section 2, a complete specification of DPmeta requires specifying the values of many hyperparameters, and we have no prior
information to make informed choices. So, we have replaced $\eta_b, S_b, \phi_1$ and $\phi_2$ with their maximum a posteriori probability
(MAP) estimates. We have followed \cite{e94} by considering $M \in \{L^{-1}, L^0, L^1, L^2\} = \{1/3, 1, 3, 9\}.$


From (2.10) in \cite{mqjh15} the prior probability of $k$ clusters is a function of $M.$ Let $p_M = (p_{1M}, p_{2M}, p_{3M})$ where $p_{kM}$
is the \emph{prior} probability of $k$ clusters with precision $M$. Then $p_{kM}$ can be calculated using the probability associated with
any partition, i.e.,

\begin{equation}\label{prior prob partition}
\frac{M^{k-1}\prod_{j = 1}^k \Gamma(L_j)}{(M+1)(M+2)\cdots   (M+L-1)}
\end{equation}
where $L_j$ is the number of surveys in cluster $j$ with $\sum_{j = 1}^k L_j = L.$ Then $p_{1/3} = (18/28, 9/28, 1/28), p_1 = (2/6, 3/6, 1/6),
p_3 = (2/20, 9/20, 9/20)$ and $p_9 = (2/110, 27/110, 81/110).$ Since $p_{1/3}$ and $p_9$ are too extreme we have emphasized $M = 1$ and
$M = 3.$ The results corresponding to $M = 1$ and $M = 3$ are very close, so only the latter are presented in Table 2, which has the same
format as Table 1.

Comparing the results from the uncertain pooling method with those from DPmeta it is apparent that, in general, there is close agreement.
For the posterior mean they are similar except in panel 3 where there is greater shrinkage for surveys 2 and 3. The results for the posterior
SD are also close except for a larger value for survey 2 in panel 3. There are only small differences in the intervals except for panel 3 where the
DPmeta intervals for surveys 2 and 3 are wider.

The small value of the SAHIE SE  seen in almost all counties limits the scope of our evaluation. Thus, we have used  modified data sets based on the original data. Here, as before, we take the CDC SE to be 0.5, 1 and 2 times the HS SE, but also take the SAHIE SE to be 2, 5 and 10 times the HS SE. Table 3, with the same format as Table 1, shows, for uncertain pooling, the results for Orange county with the CDC SE = 0.5(HS SE).
These results are typical of our analyses for a sizeable number of FL counties. Recall, though, that each analysis is based only
on the data from the specific county.
For panel 1 in Table 3, $E(\nu | \{\hat{Y}_i : i = 1,...,L\}) = 0.199, SD(\nu | \{\hat{Y}_i : i = 1,...,L\}) = 0.008$ and the 95\% credible
interval is $(0.184, 0.215).$ Inferences based on the posterior distribution of $\nu$ are inappropriate if one regards any one of the three surveys
as the ``gold standard." For example, the posterior mean of $\mu_1, \quad 0.278,$ is outside the 95\% credible interval for $\nu.$ As in Table 1,
$p(g = 1 |y) \le 0.001,$ indicating very little support for pooling all of the data.

The percent reductions in the posterior standard deviation of $\mu_1$, i.e., for SAHIE (survey 1), are 11, 26 and 44\%, corresponding to the three panels in Table 3. As the value of the SAHIE SE is increased, there is, as expected, additional pooling of the SAHIE observed proportion with the CDC observed proportion. Noting that the observed proportion for SAHIE is 0.294, the posterior means for $\mu_1$ decrease from 0.278 (panel 1) to 0.240 (panel 3). One reason for this can be seen by comparing the posterior distributions of $g,$ i.e., $\{g, p(g|y): g = 1, \ldots, 5\}$, given at the bottom of each panel. For example, $p(g =2|y)$ increases from 0.006 to 0.339 while  $p(g = 5|y)$ decreases from 0.479 to 0.253. These results show that the uncertain pooling methodology is taking proper account of the increased variability associated
with the SAHIE estimates, i.e., increasing the likelihood of pooling the data from surveys 1 and 3.

\begin{table}[h!t!b!p!]
\centering
\caption{Observed proportions, standard deviations and posterior summaries from Orange County, Florida, where CDC SE = 0.5 $\times$ HS SE using uncertain pooling.}

\setlength\tabcolsep{2pt}
\begin{tabular}{cccccc}
 \hline
 & \multicolumn{4}{c}{SAHIE SE = 2 $\times$ HS SE} &   \\
 Survey   &  ObsProp   &  PostMean    &  ObsSE   &  PostSD &  $95\%$ Cred Int  \\
1  &  0.294  & 0.278  &  0.036    & 0.032 &   (0.227, 0.352)  \\
2   &0.257  &  0.261  &  0.018   &   0.017 &  (0.226, 0.294) \\
3  & 0.179   &0.179   & 0.009    &   0.009  & (0.162, 0.197)\\
pool-all&  & 0.199 &  &0.008 &  (0.184, 0.215) \\
\multicolumn{6}{l}{\scriptsize $P(g=2|y)=0.006; P(g=3|y)=0.514; P(g=5|y)=0.479$} \\

    &   &   & &              &    \\
     & \multicolumn{4}{c}{SAHIE SE = 5 $\times$   HS SE}  &   \\
Survey   &  ObsProp   &  PostMean    &  ObsSE   &  PostSD &  $95\%$ Cred Int  \\
  1  &   0.294 & 0.251 &  0.089 & 0.066 &    (0.162, 0.417)   \\
  2  &   0.257  &  0.258 & 0.018 & 0.018  &     (0.223, 0.293    \\
  3  &  0.179   & 0.179  &0.009  & 0.009 &   (0.162, 0.197)  \\
pool-all&  & 0.195 &  &0.008 &  (0.180, 0.211) \\
\multicolumn{6}{l}{\scriptsize $P(g=2|y)=0.224; P(g=3|y)=0.468; P(g=5|y)=0.308$} \\
    &   &   & &              &    \\
     & \multicolumn{4}{c}{SAHIE SE =  10 $\times$ HS SE}  &   \\
Survey   &  ObsProp   &  PostMean    &  ObsSE   &  PostSD &  $95\%$ Cred Int  \\
  1  &   0.294 & 0.240 &  0.179 & 0.101 &    (0.059, 0.520)   \\
  2  &   0.257  &  0.257 & 0.018 & 0.018  &     (0.222, 0.292)    \\
  3  &   0.179   & 0.179  & 0.009  & 0.009 &   (0.162, 0.197)  \\
pool-all&  & 0.195 &  & 0.008 &  (0.179, 0.211) \\
\multicolumn{6}{l}{\scriptsize $P(g=2|y)=0.339; P(g=3|y)=0.408; P(g=5|y)=0.253$} \\
     \hline
\end{tabular}

\end{table}

\begin{table}[h!t!b!p!]
\centering
\caption{Observed proportions,  standard errors and posterior summaries from Orange County, Florida using DPmeta.}

\setlength\tabcolsep{4pt}
\begin{tabular}{cccccc}
 \hline
Survey & ObsProp   &PostMean & ObsSE &  PostSD &   $95\%$ Cred Interval \\
     &  \multicolumn{5}{c}{SAHIE SE =  2 $\times$ HS SE}   \\
1  &  0.294     &  0.262 &0.036 &  0.021 &     (0.202, 0.297) \\
2   &0.257     &   0.263 &0.018 &  0.018 &    (0.222, 0.296)  \\
3  & 0.179      & 0.180 & 0.009&     0.009  &   (0.162, 0.199)\\
   & &  &   &  &     \\
     &  \multicolumn{5}{c}{SAHIE SE =  5 $\times$ HS SE}  \\
  1  &   0.294 &  0.226 & 0.089 &  0.040 &       (0.168, 0.290)    \\
  2  &   0.257  &    0.246 &0.018 &  0.023  &        (0.186, 0.291)  \\
  3  &  0.179   &    0.182 &0.009 & 0.011 &       (0.163, 0.205) \\
   & &  &   &  &    \\
   &    \multicolumn{5}{c}{SAHIE SE =  10 $\times$ HS SE}   \\
  1  &   0.294 &   0.217 &0.179 &   0.044 &        (0.166, 0.288)  \\
  2  &   0.257  &   0.243 &0.018&   0.031  &        (0.185, 0.290)    \\
  3  &  0.179   &   0.183  &0.009 &  0.011&      (0.162, 0.205)   \\
     \hline
\end{tabular}

\end{table}

Comparing the results from the uncertain pooling method in Table 3 with those from DPmeta in Table 4 it is apparent that there
are greater differences than those seen in Tables 1 and 2. For the posterior means it is notable that for survey 1 the posterior mean from
DPmeta is somewhat smaller than that from uncertain pooling. This reflects greater pooling of the data from survey 1 with
that from survey 3. For surveys 2 and 3 the two sets of posterior means are similar.
The most notable difference is for
the posterior SDs where, for survey 1 (panels 2 and 3) the values are very much smaller from DPmeta than from uncertain pooling.
For the other seven cases, the two sets of posterior SDs are similar.
Correspondingly, for survey 1 the intervals from DPmeta are much shorter than those from uncertain pooling while those for
surveys 2 and 3 are only a little wider. From Table 4 and for survey 1 the percent reductions in the posterior SD (relative to
the ObsSEs) are (42\%, 55\%, 75\%). There are increases, though, for survey 2 in panels 2 and 3.

\subsection{Results from simulation study}
To evaluate properties such as bias and coverage of the credible interval we have carried out a simulation study based
on several modifications of the Orange county data. Specifically, we generate $\{\hat{Y}_i : i = 1,2,3\}$ from

\begin{eqnarray}\label{simulation model}
  \hat{Y}_1 & \sim & N(\psi_1, V_1) \nonumber \\
  \hat{Y}_2 & \sim & N(\psi_1, V_2) \nonumber \\
  \hat{Y}_3 & \sim & N(\psi_2 + \Delta, V_2)
\end{eqnarray}
where $\psi_1$ and $\psi_2$ are from the third panel of Table 3; $\psi_1$ is the average of the observed proportions from surveys
1 and 2 while $\psi_2$ is the observed proportion from survey 3 (CDC). Also, we took $V_1$ to be much larger than $V_2.$ These
choices were made to represent a common situation where survey 1 is a probability sample, with relatively large sample variance
while surveys 2 and 3 are non-probability samples with much smaller sample variances.
Finally, $\Delta \in \{0,4(0.0193)= 0.0772, 8(0.0193)= 0.1544\}$.

Table 5 gives the values of $\psi_1, \psi_2, V_1$ and $V_2$ in the footnote. There are three rows, corresponding to $\Delta = 0,
0.0772, 0.1544$. In each row there are the medians over 500 replications of $\{p(g|y): g = 1,...,5\}, \{E(\mu_i | y) : i = 1,2,3\}$
and $\{SD(\mu_i | y): i =1,2,3\}$, together with the estimated coverages.

\begin{table}[h!t!b!p!]
\centering
\caption{Simulation results from 500 replications of (\ref{simulation model});
parameter values are in the footnote.}
\

\setlength\tabcolsep{2pt}
\begin{tabular}{l|ccccc|ccc|ccc|ccc}
 Inc Size & \multicolumn{5}{c|}{$P(g | \text{data} ) $} & \multicolumn{3}{c|}{Coverage}  &  \multicolumn{3}{c|}{PostMean}  & \multicolumn{3}{c}{PostSD}  \\
&$g : 1$  &2 &3 & 4&5 & $i:1$ & 2&3 & $i:1$ & 2&3 & $i:1$ &2 &3  \\
 \hline
$\Delta =0$          &  0  & 0.148  &  0.462    & 0 & 0.332 & 0.973 & 0.958 & 0.960 & 0.262 &0.275 & 0.179 & 0.050 &0.006 & 0.006  \\
$\Delta =0.0772$    &  0.032  & 0.292  &  0.303    & 0.033 & 0.249 & 0.984 & 0.941 & 0.939 & 0.269 &0.275 & 0.257 & 0.039 &0.006 & 0.006  \\
$\Delta =0.1544$   &  0  & 0.280  &  0.401    & 0 & 0.300 & 0.971 & 0.958& 0.952 & 0.292 &0.275 & 0.333 & 0.044 &0.006 & 0.006  \\
\hline
\multicolumn{15}{l}\scriptsize $\psi_1 = 0.276, \psi_2 = 0.179, V_1=0.06^2, V_2=0.006^2$
\end{tabular}

\end{table}

The principal findings are: (a) the medians of the posterior means are close to the values used to generate the data, i.e., $\psi_1$ and $\psi_2$,
(b) the coverages are close to the nominal 95\%, and (c) there are significant reductions in the posterior standard deviation for survey 1 (SAHIE), i.e., 16.7\%, 35.0\% and 26.7\% corresponding to $\Delta = 0, 0.0772$ and $0.1544.$ There are no reductions in the posterior standard deviations for surveys 2 and 3.

For $\Delta = 0.1544$ note that $p(g =2|y) = 0.280$ while $p(g = 4 | y) = 0.$ That is, we pool data from surveys 1(SAHIE) and 3(CDC),
$\{g = 2\}$, because of the relatively large SE for survey 1(SAHIE). However, we do not pool data from surveys 2(HS) and 3(CDC),
$\{g =4\}$, because of the relatively small SEs for survey 2(HS) and survey 3 (CDC). Of course, we pool data from surveys 1 and 2, $\{g = 3\},$
because they have the same mean, $\psi_1.$

\section{Discussion and Summary}

With reduced response rates and diminished resources there is considerable interest in combining data from several sources such as
sample surveys and administrative data. Currently there is special interest when the sources include non-probability surveys. Appropriate methodology is required to ensure satisfactory inferences since the target populations and data acquisition methods may be quite different.

There are many situations where it may be beneficial to combine such data, as shown in the review paper by \cite{lar17}. Here, we have investigated the case where the analyst has  only summary statistics from each of the
sources, and where one can think of one source, $r$, as the single best source for inference. While it is often beneficial to use the data from
related sources to improve inferences from $r$, it is essential that the data that are combined be concordant with the
data from $r$. The methodology in this paper can also be used in settings where the data are not limited to summary statistics
and inferential objectives and models are more
complex. As seen in this paper, failure to consider biases due to pooling ``unlike" data may lead to poor inference. Using analytical expressions and examples we have shown that both the uncertain pooling and DPM methods provide appropriate
inferences. However, our analyses based on uncertain pooling are fully Bayes while those from DPmeta are empirical Bayes --
due to the need to specify values for many hyperparameters. Moreover, the uncertain pooling method provides additional information in
the form of the posterior probabilities for the partitions, $g.$

The methods can be implemented. For DPmeta there is an R package, DPpackage \citep{jhqmr11} while an R package is being developed for the
uncertain pooling method. When completed it will be submitted to The Comprehensive R Archive Network. It contains functions that allow
Bayesian analyses of the type described in this paper (a)with user-supplied point estimates and associated variances, or (b)with 
binomial data, cases ($y$) and total counts ($n$). Case (b) provides an analysis based on the logit transformation of the sample proportion.
We have implemented the latter when there are eleven surveys.

Making inference for the sample variances, $V_1,...,V_L$ is a very challenging problem. \cite{p17} provides an extensive discussion
of methods that have been proposed. Of particular interest are solutions proposed by \cite{yc06}, \cite{stk17} and \cite{p17}.
However, these solutions are posed in the context of small area inference, not
when the objective is combining data from surveys and related sources.

While the discussion below is in the context of extending the DPM method (Section 2.2), the ideas are also relevant for the uncertain pooling
method (Section 2.1). \cite{p17} augments the DPM model in Section 2.2 with

\begin{equation}\label{sample variance}
\delta_i S^2_i \stackrel{ind}{\sim} V_i \chi^2_{\delta_i}, \quad i = 1,...,L
\end{equation}
and
\begin{equation}\label{pop variance}
V_i^{-1} \stackrel{iid}{\sim} \text{Gamma} (a_1, b_1).
\end{equation}
where $S^2_i$ is the sampling variance and $\delta_i$ is a measure of the degrees of freedom.

As noted by \cite{p17} the assumption of the $\chi^2$ distribution in (\ref{sample variance}) is questionable, surely so  when there is a complex survey design.  With only a single sample one cannot verify the \emph{sampling distribution} of
$S^2_i$, a point also made by \cite{p17} on page 731. Moreover, in survey sampling the form of $S^2_i$ is likely to be
a complex function of the values of the variable of interest, $Y,$ and survey weights. Thus, it is unlikely that the
distribution of the observed $Y$'s can be used to infer a reasonable approximation for the distribution of $S^2_i.$

The assumption of constant population parameters in (\ref{pop variance}) is problematic for our case, i.e., combining data.
We expect
considerable differences among the surveys, e.g., for a collection of probability and non-probability samples. \cite{yc06}
generalize by replacing (\ref{pop variance}) with
\begin{equation}
V_i \stackrel {ind}{\sim} \text{Inverse Gamma} (a_i, b_i).
\end{equation}
This requires values for $(a_i, b_i)$, a difficult choice without
prior information. Moreover, \cite{ge06} shows that selecting both $a_i$ and $b_i$ very small, a natural choice
(and one made by \cite{yc06}), may lead to poor inferences. \cite{stk17} provide an alternative to \cite{yc06} by assuming
\begin{equation}
V_i \stackrel {ind}{\sim} \text{Inverse Gamma} (a_i, b_i\gamma),
\end{equation}
with a prior on $\gamma$, but this, too,
requires specifying values for $a_i$ and $b_i.$
Clearly, making better inference for the sample variances is an important topic for future research.

There has been an increased interest in making inference for small subpopulations, i.e., ``small area" inference, when there are
several data sources; see, e.g., \cite{mea11} and \cite{nbb14}. While further research is needed to extend the uncertain pooling methodology
to this case the approach is clear. Let $j$ denote a small area, e.g., a US county, and $i$ denote a data source where $j = 1,...,J$
and $i = 1,..., L$. As above $g$ denotes a generic partition with generic subset $S_k(g)$ for $k = 1,...,d(g)$.
Define $\mathcal{G} = \{(ij): j = 1,...,J; i = 1,...,L\}.$ Then for fixed $g$, $S_k(g)$ is a subset of $\mathcal{G}$ with
$S_k(g) \bigcap S_m(g) = \emptyset$ for $k \neq m$ and $\bigcup_{k=1}^{d(g)}S_k(g) = \mathcal{G}$.
For example, let $J =2$ and $L =3$. Then each partition is a collection of the disjoint
subsets of $\mathcal{G} = \{(11), (12), (21), (22), (31), (32)\}$ whose union is $\mathcal{G}.$ By analogy with the
discussion in Section 2, there would be a single best source for each small area, identified as $(j,i(j))$ for some $i$ in small
area $j.$

Then the following model, analogous to the one in Section 2, is

\begin{equation}
\hat{Y}_{ij} \stackrel{ind}{\sim} N(\mu_{ij}, V_{ij}).
\end{equation}

By analogy with (\ref{first stage prior})
\begin{equation}
 \mu_{ij} \stackrel{ind}{\sim} N(\nu_k(g), \delta_k^2(g)),\qquad ij \in S_k(g).
\end{equation}

If the same (limit) assumptions are made about the $\gamma^2_k(g)$ and $\delta_k^2(g) = \delta^2$,
the expressions for posterior inference for the $\mu_{ij}$ will be the same as in Section 2.
However, the assumption of constant $\delta^2$ may not be reasonable. Since there will be a very
large number of partitions computation will be challenging, especially since it is expected that many
$p(g|y)$ will be very small.

The premise of our work is that one should include the possibility that the parameters associated with different surveys may not be exchangeable. (With a probability sample and several non-probability samples this may be especially important.) Similarly, it is natural to generalize so that the parameters associated with the small areas are not assumed to be exchangeable. However, if exchangeability across both surveys and small areas can be assumed, the model in Section 2.1 of \cite{kpk15} (possibly modified to accommodate Bayesian inference) should be easier to implement.

Although the very large number of partitions of $\mathcal{G}$ may pose an obstacle to implementation, one may
be able to apply DPmeta when there are data from a set of small areas and several data sources. One problem is
the specification in DPmeta of a common distribution for the $\mu_{ij}$, i.e., over small areas
and surveys, which is unlikely to be appropriate. The possibilities
include an ANOVA model (Section 4.4.2) or nested model (Section 7.3.1) of \cite{mqjh15}, although
the ANOVA model has no interaction terms and our model is a cross-classified one.

Future research should include making inference for the sample variances, as noted above. Also, we need
improved methodology to handle the extension to small area inference when there are data from several surveys.
In some cases one may be able to simplify the model for the $\mu_{ij}.$ Using a grid-based method for sampling
$g$ and $\delta^2$ is difficult to implement when $G$ is extremely large. So, using a standard MCMC approach, possibly with an
informative prior on $g$, may be a better way to make inference. For example, see \cite{ddt17}.

Other approaches could also be explored.
For example,
\cite{pks17} suggest a different approach for combining data from two surveys. Here, there are covariates, X, observed
in each survey while $Y_1,$ the study variable of interest, is observed only in survey $1$ and $Y_2$ is observed only in
survey $2$. Inference for the population mean of $Y_1$ is desired, given data from both surveys. The densities that
they use are $f_1(Y_1 | X, \theta_1), f_2(Y_2| X, Y_1, \theta_2)$ and, for identifiability, it is assumed that
$f_2(Y_2 | X, Y_1) = f_2(Y_2 | Y_1).$
For a Bayesian analysis an extension to more than two surveys would be needed, together with specification of appropriate
prior distributions for the parameters. It does not seem to be straightforward to model the distribution of $Y_2$ given
$Y_1.$

\section{Acknowledgment}

\emph{The authors are grateful to the reviewers whose extensive comments have resulted in a paper  that has greater focus and increased breadth. They also appreciate research allocation grants from ACCESS's Pittsburgh Supercomputing Center.}

\section{References}
\renewcommand*{\refname}{}

\section{Appendix: Small Area Health Insurance Estimates (SAHIE) Program}

The following summary paraphrases relevant parts of \cite{cb21}. To avoid distortion of the authors' meaning we have retained the first-person text.

The SAHIE program produces model-based estimates of health insurance coverage for demographic groups within counties and states. We
publish county estimates by sex, age and income. The income groups are defined by the income-to-poverty ratio (IPR) - the ratio of
family income to the appropriate federal poverty level.

For estimation, SAHIE uses models that combine survey data from the American Community Survey (ACS) with administrative records data
and Census data. The models are ``area-level" models because we use survey estimates and administrative data at certain levels of
aggregation, rather than individual survey and administrative records. Our modeling approach is similar to that of common models
developed for small area estimation, but with additional complexities.

The published estimates are based on aggregates of modeled demographic groups. For counties, we model at a base level defined by
age, sex and income groups.

We use estimates from the Census Bureau's Population Estimates Program for the population in groups defined for county by age and sex.
We treat these populations as known. Within each of these groups, the number with health insurance coverage in any of the income
categories is given by that population multiplied by two unknown proportions to be estimated: the proportion in the income category
and the proportion insured within that income category. The models have two largely distinct parts - an ``income part" and an ``insurance
part" -- that correspond to these proportions. We use survey estimates of the proportions in the income groups and of the proportions
insured within those groups. We assume these survey estimates are unbiased and follow known distributions. We also assume functional
forms for the variances of the survey estimates that involve parameters that are estimated. We treat supplemental variables that predict
one or both of unknown income and insurance proportions in one of two ways:

Some of these variables are used as fixed predictors in a
regression model. There is a regression component in both the income and insurance parts of the model. In each case, a transformation
of the proportion is predicted by a linear combination of fixed predictors. Some of these predictors are categorical variables that
define the demographic groups we model. Others are continuous. The continuous fixed predictors include variables regarding employment,
educational attainment, and demographic population.

We also utilize random continuous predictors, which include data from 5-year ACS, Internal Revenue Service, Supplmental
Nutrition Assistance Program, and Medicaid/Children's Health Insurance Program. These are not fixed predictors in the model.
Instead, we treat them as random, in a way similar to survey estimates, but not as unbiased estimators of the numbers. Instead,
we assume that their expectations are linear functions of the number in an income group or the number insured within an income
group. We typically assume they are normally distributed with variances that depend on unknown parameters.

We formulate the model in a Bayesian framework and report the posterior means as the point estimates. We use the posterior means
and variances together with a normal approximation to calculate symmetric 90-percent confidence intervals, and report their half-widths
as the margins of error.

We control the estimates to be consistent with specified national totals.


\begin{thebibliography}{9}
\bibitem[Malec and Sedransk(1992)]{mas92} Malec, D., and Sedransk, J.  (1992), ``Bayesian methodology for combining the results from different experiments when the specifications for pooling are uncertain,"  \emph{Biometrika}, 79(3), 593--601.
\end{thebibliography}
\end{document}